\documentclass[preprint]{elsarticle}

\usepackage{lineno,hyperref}
\usepackage{amsmath} 

\journal{Journal of \LaTeX\ Templates}

\begin{document}

\begin{frontmatter}

\title{Temperature dependence of mobility of conducting polymer polyaniline with secondary dopant}

\author[mymainaddress]{K. Yamada}
\ead{yamada@phys.kyushu-u.ac.jp}

\author[mymainaddress]{B. Shinozaki}
\author[mymainaddress]{T. Narikiyo}
\author[mymainaddress]{Y. Takigawa}
\address[mymainaddress]{Department of Physics, Kyushu University, Fukuoka, Japan}

\author[mysecondaryaddress]{N. Kuroda}
\author[mysecondaryaddress]{T. Bandou}
\author[mysecondaryaddress]{H. Nakamura}
\address[mysecondaryaddress]{Advanced Technology Research Laboratory, Idemitu Kosan Co. Ltd., Sodegaura, Japan}

\cortext[mycorrespondingauthor]{Corresponding author}

\begin{abstract}

The conductivity $\sigma$ and carrier density $n$ of the conducting polymer polyaniline were investigated by changing the concentration $x$ of a secondary dopant, meta-cresol. We found that $\sigma$ changes by four orders of magnitude within the $x$-range of 1--10\%, while $n$, as estimated from the Hall measurements, shows a weak dependence on $x$ in the region of 2 \% $<x<50$ \%. These results suggest that $\sigma$ can be enhanced by the change in the mobility $\mu$. We analyzed the temperature dependence of $\mu$ by not only the combination of two different types of scattering mechanism, but also by the polaron hopping model. The experimental data of $\mu(T)$ can be explained well by the latter model with reasonable fitting parameter values of a small-polaron binding energy and a longitudinal optical phonon frequency.
\end{abstract}


\end{frontmatter}


\section{Introduction}
Conducting polymers can be solution-processed and are useful for low-cost solar cells and printed circuits\cite{Heegerbook}. To improve devices for applications and to understand the conduction mechanism, investigations of the mobility $\mu$ and carrier density $n$ of conducting polymers are very important. Although there has been research on the mobility of the organic surface region in contact with the insulator by the Hall effect and field effect for organic field-effect transistors, the temperature dependence of the Hall mobility of uniform conducting polymers has not been researched in-depth\cite{Yamagishi10PRB, Uemura12PRB, Xinge11PRB, Lee12NM, Senanayak15PRB}. For simple conducting-polymer monolayer films, the temperature dependence of $\mu$ and $n$ has not been independently reported, although the electrical conductivity $\sigma$ has been extensively investigated from the viewpoint of the metal-insulator transition\cite{Long04PHB}.

Among various conductive polymers, polyaniline (PANI) is well-investigated\cite{Heegerbook}. The metallic transport properties of PANI have also been reported\cite{Lee00nat}. Doping first with a substance such as a sulfosuccinic ether acid increases the hole carrier concentration and enables PANI to dissolve in a solvent\cite{Heegerbook, Kuramoto97POL, Kuramoto00SM}. Furthermore, in terms of adaptation for a secondary dopant, it is known that the conductivity of PANI is increased up to $\sim$10000 times than without a secondary dopant\cite{Heegerbook, MacDiarmid95SM, MacDiarmid94SM}. MacDiarmid and Epstein reported that the effects of secondary doping are based primarily on a change in molecular conformation of PANI from a compact coil to an expanded coil that occurs during the secondary doping process\cite{MacDiarmid94SM,MacDiarmid95SM}.
Contrarily, optical studies of PANI have been reported \cite{Heegerbook, Stafstrom87PRL, Kohlman96PRL}. These optical studies suggest that the carriers are polarons with a heavy effective mass. However, Lee~{\it et al.~} reported that the optical conductivity of high-conductivity PANI camphor sulfonic acid can be successfully described by the simple Drude model without incorporating contributions from the disorder-induced localization theory\cite{Heegerbook, Lee00nat}. To date, however, systematic investigations of $\mu(T)$ from the viewpoint of scattering mechanisms have not been extensively undertaken.
We study the origin of the significant enhancement of the conductivity caused by the secondary dopant, meta-cresol(MCR), by measuring the dependences of $\sigma$ and $n$ on the dopant concentration. Further, we discuss in detail the scattering mechanism of carriers of typical films in the crossover region between high and low conductivities for a PANI film prepared with a controlled MCR concentration.

\section{Sample preparation and experimental procedure}
To prepare the PANI films, we dissolved 3.6 g of the first dopant, dioctyl sulfosuccinate sodium salt, and 3.74 g aniline in 100 mL toluene. Next, we added 300 mL 1 N HCl to the mixture during cooling. Further, by adding a mixture of 5.36 g ammonium persulfate (APS) and 100 mL 1 N HCl, the aniline could polymerize, as the polymerization of PANI was carried out by chemical oxidation using an APS oxidant in an aqueous solution. Dark green PANI suspensions were obtained from this procedure. We removed the water and volatile components from the above mixture to obtain the protonated PANI complex solid. The solid PANI was dissolved using a mixture of toluene and the secondary dopant, MCR, as the solvent, with a change in the ratio $x$(\%) of MCR to the PANI complex solid, where $x$ is defined by $x$=100 (mass of MCR) / (mass of PANI complex solid). By casting the PANI melted in the solvents on a polypropylene film with an area of 10 mm$^2$, followed by drying for 18 h at 300 K and for 1 h at $\sim350$ K in a grow box filled with N$_2$ gas, we obtained films with various resistivities, the film thickness for all specimens was maintained at $15 \mathrm{\mu m} $.

This measurement was performed by the four-terminal method with Hall bar geometry. For the electrodes, we deposited thin Au contacts on the PANI films. An external magnetic field $H$ was perpendicularly applied to the film surface, of up to $H=1.35$~T. The equipment used for the Hall measurement was reported in a previous paper\cite{Yamada12JP}. We used the Keithley 6517B and 6514 electrometers for the DC voltage measurements, as well as the Keithley 220 programmable current source. After degassing at $T=313$~K in a vacuum for 1 h to stabilize the resistance, we measured the temperature dependence of $\sigma$ and $n$ in the temperature range of 80~K~$<T<$~300~K.

\section{Experimental results and discussion}

\begin{figure}
\includegraphics[width=120mm]{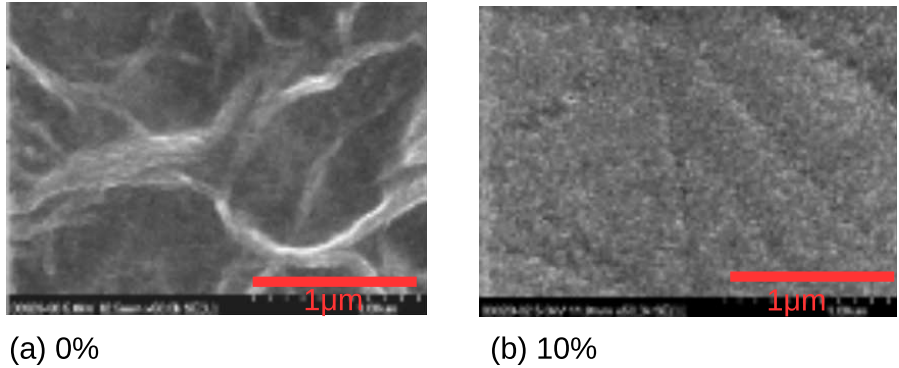}
\caption{\label{fig:SEM} 
Images of films (a) $x=0$\% and (b) 10\% obtained by scanning electron microscopy.
 }
\end{figure}
\begin{figure}
\includegraphics[width=30mm]{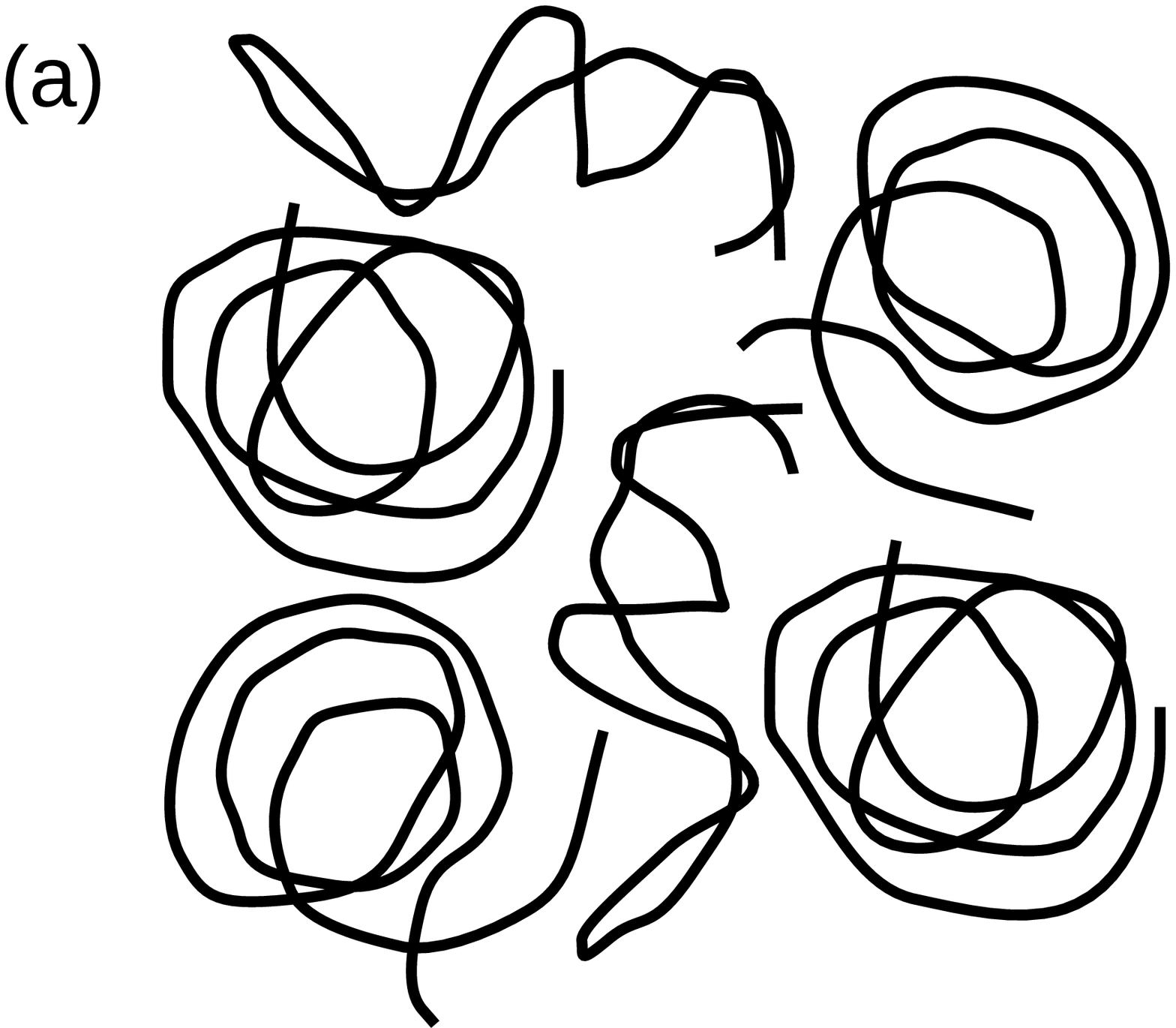}
\includegraphics[width=30mm]{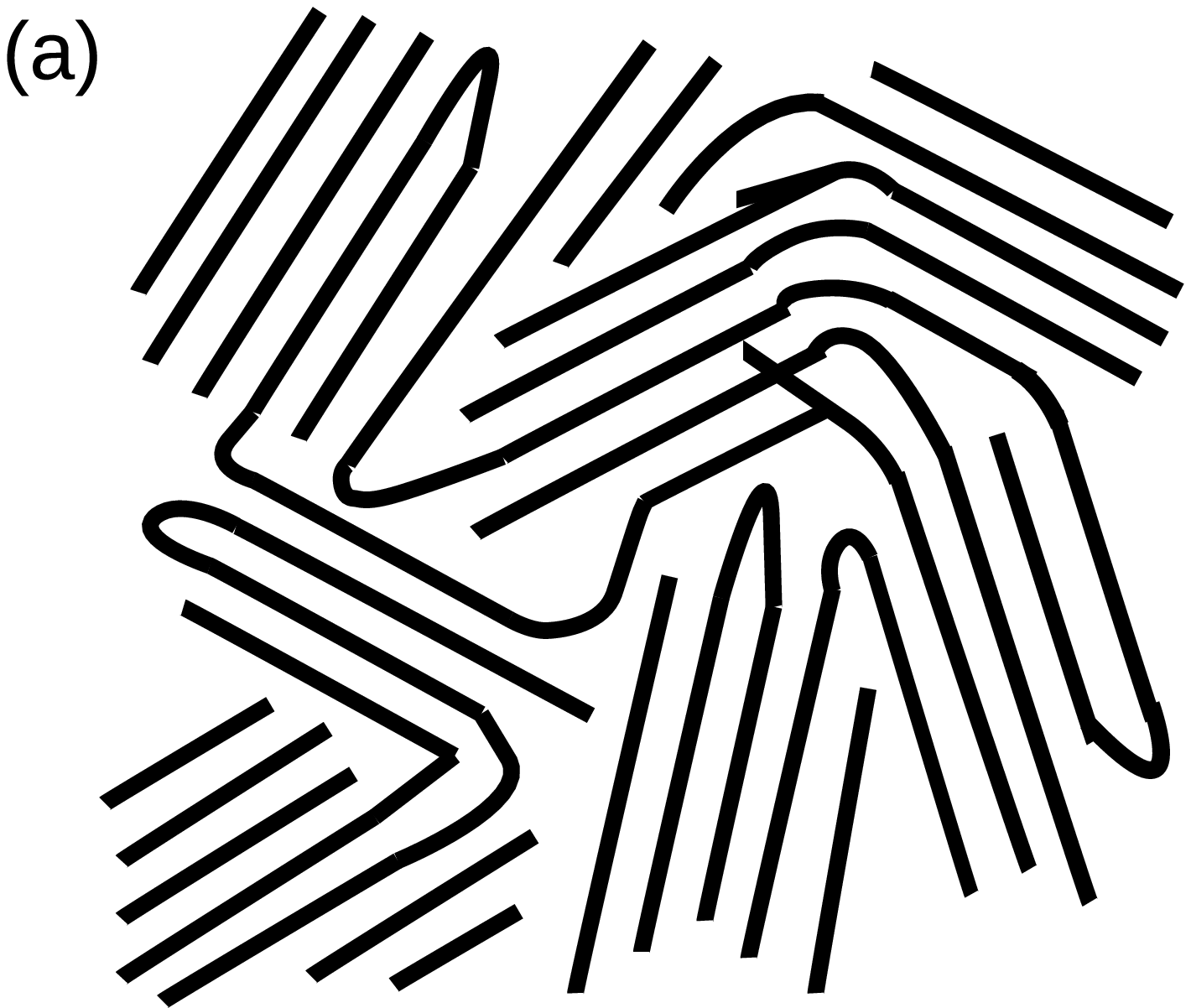}
\caption{\label{fig:cresol_model} 
Schematics of conformations of polymers in Fig. \ref{fig:SEM}. (a) bending structure of polymers caused by attraction between ionized segments of the polymer chain for the film without MCR and (b) extended structure polymer chain for the film with MCR.}
\end{figure}

Fig.~\ref{fig:SEM} shows the scanning electron microscope images for the films with (a) $x$ = 0 and (b) $x$ = 10\%. A considerable difference can be seen in the conformation between the two films. Fig.~\ref{fig:cresol_model} schematically shows the structures of the films (a) without MCR and (b) with MCR. For film (a), prepared from only the nonpolar solvent toluene without MCR, it is expected that the conformation of polymers in the film take bending and/or bundle-like structures, because the additional negative ions of the first dopant, between the positively charged sites on the polymer chain, tend to decrease the repulsive interaction and improve the compact coils of the polymer\cite{MacDiarmid94SM, MacDiarmid95SM}. As the ratio of the more-polar MCR in the solvent is increased, the interaction between the doped polymer chain and solvent also increases\cite{MacDiarmid94SM, MacDiarmid95SM}. The solvation of the ions increased, resulting in their spreading further apart and causing an expansion of the initial compact coil conformation of the PANI chain, as shown in (b). Then, the conformation of the films with MCR seems to be homogeneous compared with that of the films without MCR, that is, $x = 0$. It has been reported that MCR doping increases the crystallinity, given by the sharp peaks observed in the X-ray diffraction spectrum of PANI with MCR\cite{Varm12PI}. Although carriers in the film with the extended polymers are delocalized, it is expected that the carriers in a disordered bending polymer, that is, the wavefunction, are localized by Anderson localization. 
This localization causes low conductivity in disordered systems.

\begin{figure}
\includegraphics[width=80mm]{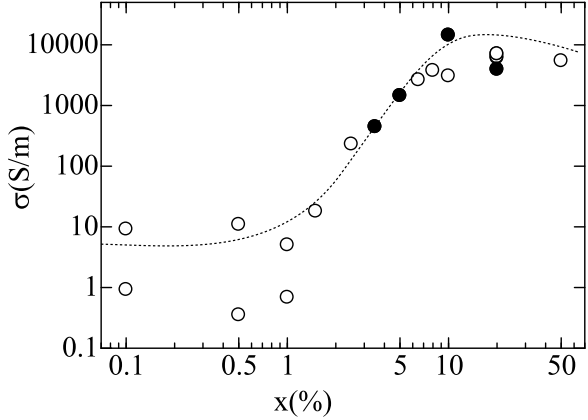}
\caption{\label{fig:conductivity} 
The m-cresol concentration dependence of conductivity. The $\sigma(T)$ for the films marked by the closed circle $(\bullet)$ is shown in Fig.~\ref{fig:resisivity-temp}. The dashed line provides a guide for the eye.}
\end{figure}

The $x$-dependence of $\sigma$ at room temperature for PANI films over a wide range of $x$ values is shown in Fig.~\ref{fig:conductivity}.
It was found that films with different magnitudes of $\sigma$ showed the same value of $x$.
This may be due to the adding effect during different waiting times, from sample preparation to measurement, and the uncontrollable conditions of film preparation.
The aging effect was expected to be dominant for low-$\sigma$ films when considering the fact the instability of the conductivity is larger for less-conductive films.
Therefore, for films in an especially low $x$-region, $x<1$\%, we considered that higher values of $\sigma$ are reasonable, as shown by dotted line.

\begin{figure}
\includegraphics[width=120mm]{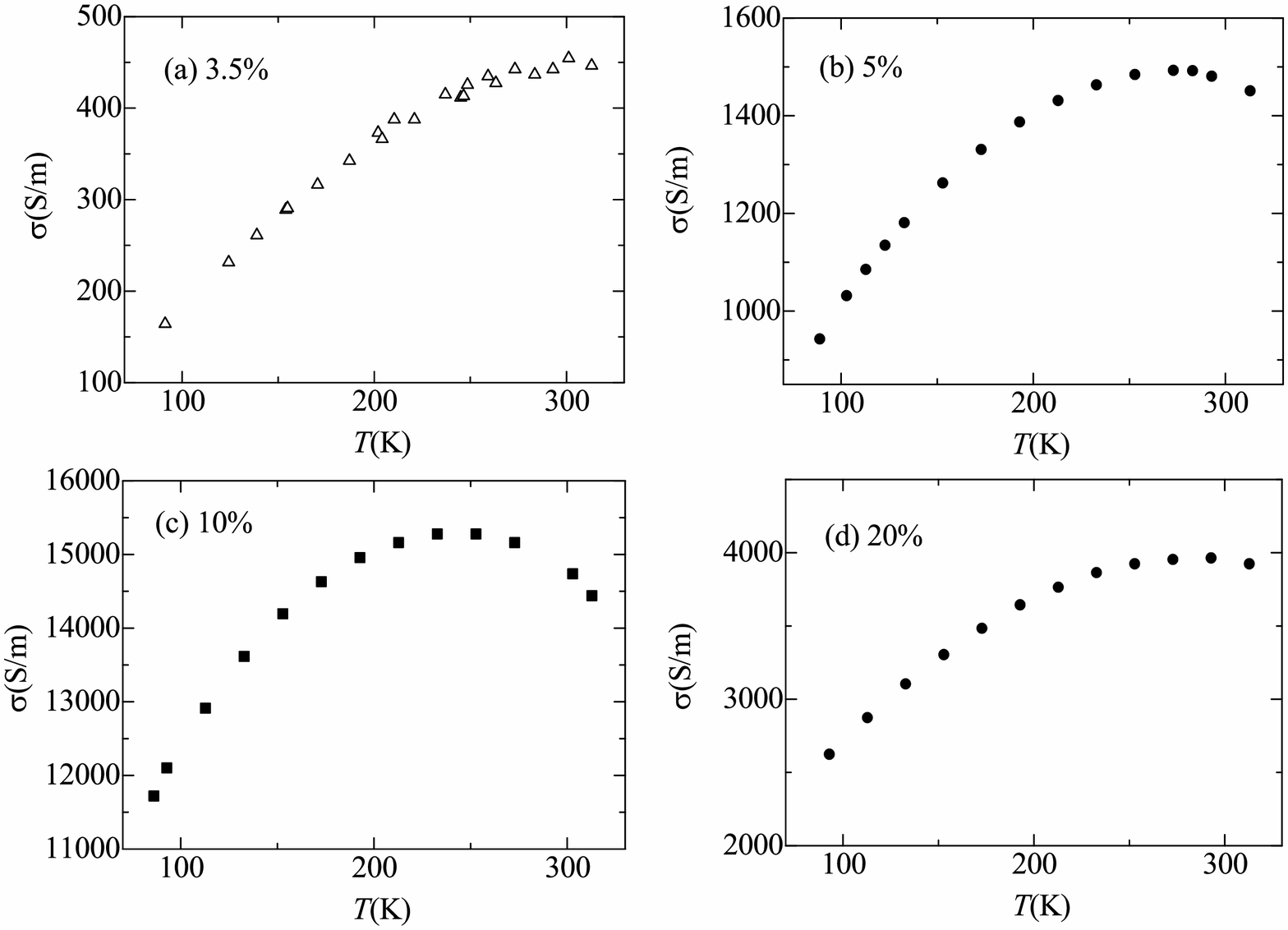}
\caption{\label{fig:resisivity-temp} 
The temperature dependence of the conductivity for the four typical films with different values of $x$: (a)3.5\%, (b)5\%, (c)10\%, and (d)20\%.}
\end{figure}

\begin{figure}
\includegraphics[width=120mm]{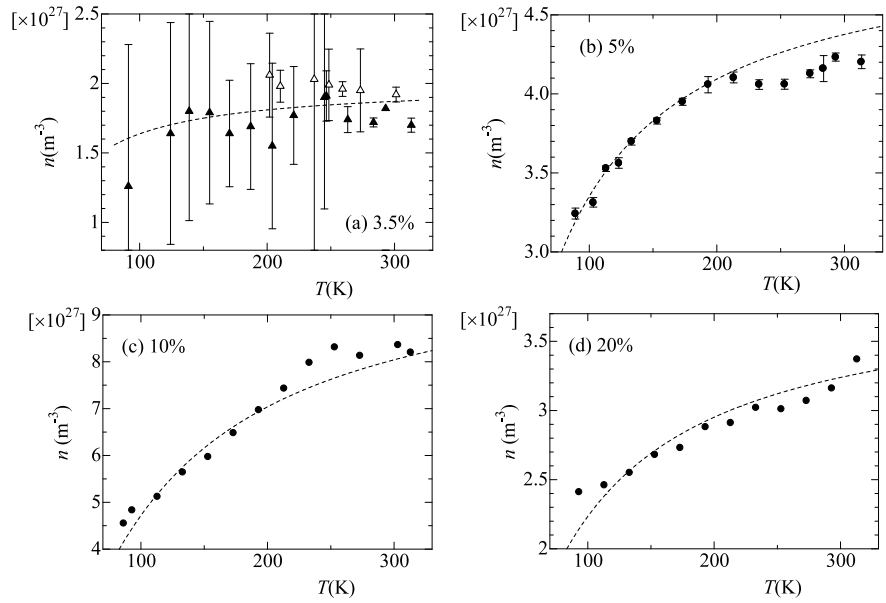}
\caption{\label{fig:n_fit}
The temperature dependences of the carrier density in films with m-cresol concentration (a) 3.5\%, (b) 5\%, (c) 10\%, and (d) 20\%. The error bars in (a) and (b) show the standard error of the mean (SEM). The SEM of (c) and (d) is similar to that of (b). The closed triangles show the data which were measured one week after that of the open triangles. The dotted lines were calculated from Eq.~\eqref{eq:density_activation} to fit the experimental data.}
\end{figure}

\begin{figure}
\includegraphics[width=120mm]{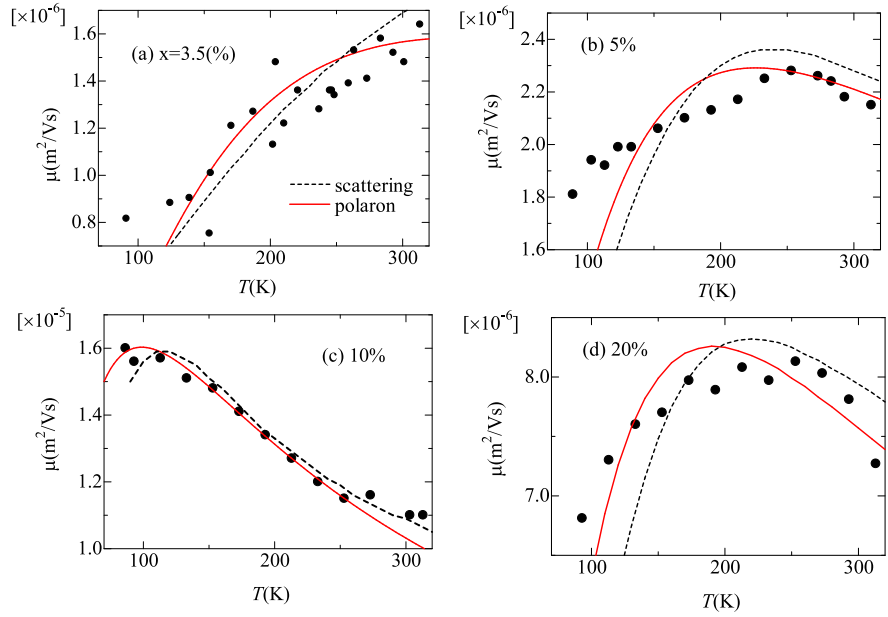}
\caption{\label{fig:mb_fit}
The temperature dependence of the Hall mobility for m-cresol concentration (a) 3.5\%, (b) 5\%, (c) 10\%, and (d) 20\%. The black dash and red sold lines were calculated from Eqs.~\eqref{eq:total_mob} and ~\eqref{eq:polaron_mob} to fit the experimental data, respectively.}
\end{figure}

A large enhancement in $\sigma$ was observed in the region of $1\%<x<10\%$ as $x$ increased. To find the reason for this strong dependence of $\sigma$ on $x$, from the viewpoint of the scattering mechanism, it was necessary to first investigate the $x$-dependence of $n$.
We succeeded in measuring the value of $n$ for films with $x \geq 2.5\%$ at room temperature. Particularly, for the four films shown by closed marks, the $T$-dependence of $\sigma$, $n$, and the mobility $\mu$ is shown in detail in Figs.~\ref{fig:resisivity-temp}, \ref{fig:n_fit}, and \ref{fig:mb_fit}, respectively.

\begin{figure}
\includegraphics[width=80mm]{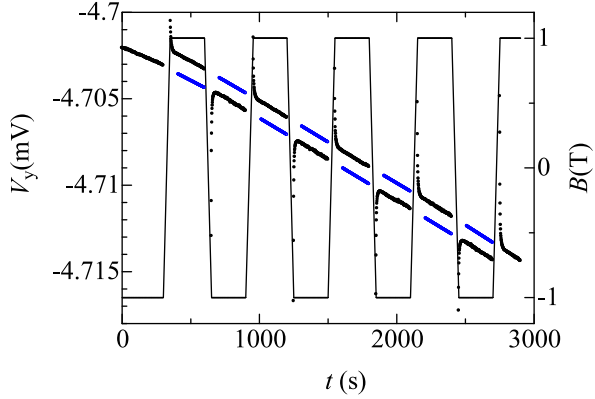}
\caption{\label{fig:typical} 
Time evolution of the Hall voltage $V_\mathrm{y}$ of the films with $x=10$\% at $T = 313$~K and $B=\pm1$. The black dots and blue lines show the experimental data and interpolation lines of Hall voltage, respectively.
}
\end{figure}

As the resistance showed a gradual change with time, we measured the Hall voltage $V_\mathrm{y}$ with a high accuracy over a sufficient time. In Fig.~\ref{fig:typical}, typical data of the time-evolution of the $V_\mathrm{y}$ are shown for the film $x=10$\% at $T = 313$~K and $B=\pm 1$~T with a period of 600 s, indicated by the nearly rectangular function (right vertical axis). The black and blue lines show the experimental data and expected values of $V_\mathrm{y}$, respectively; the latter values were obtained by interpolating the experimental data of $V_\mathrm{y}$ between the before and after intervals. We determined the voltage difference $\Delta V_\mathrm{y}=V_\mathrm{y}(B)-V_\mathrm{y}(-B)$ between the experimental data and interpolation values.

\begin{figure}
\includegraphics[width=80mm]{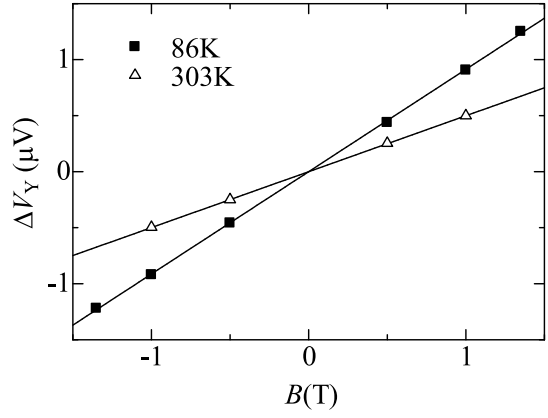}
\caption{\label{fig:typical_VB} 
The magnetic field dependence of $\Delta V_\mathrm{y} = V_\mathrm{y}(B)-V_\mathrm{y}(0)$ for the film $x=10$ \% at $T$ = 303~K and 86~K.}
\end{figure}

To clarify the linear relationship between $V_\mathrm{y}$ and $B$, data for two typical temperatures are shown in Fig.~\ref{fig:typical_VB}, where the $\Delta V_\mathrm{y}= V_\mathrm{y}(B)-V_\mathrm{y}(0)$ measurements were taken between the two values $B=0$ and $\pm B$.
As the magnetic field $B$ was reversed, the voltage $\Delta V_\mathrm{y}$ also reversed. An accurate linear relationship between $\Delta V_\mathrm{y}$ and $B$ was obtained (The coefficient of determination $R^2 = 0.9998$ of the least squares method was obtained for $x=10$ \% at $T$ = 303~K). The linear relationship confirms the high accuracy of the experimental data.
Assuming the free electron model, the carrier density $n$ can be obtained from the relationship: $ R_\mathrm{H} = E_\mathrm{y} /( B_\mathrm{z} j_\mathrm{x} )=- 1/(ne)$, where $R_\mathrm{H}$, $E_\mathrm{y}, j_\mathrm{x}$, and $B_\mathrm{z}$ are the Hall coefficient, electrical field in the y-axis direction, current density in the x-axis direction, and magnetic field in the z-axis direction, respectively.

\begin{figure}
\includegraphics[width=80mm]{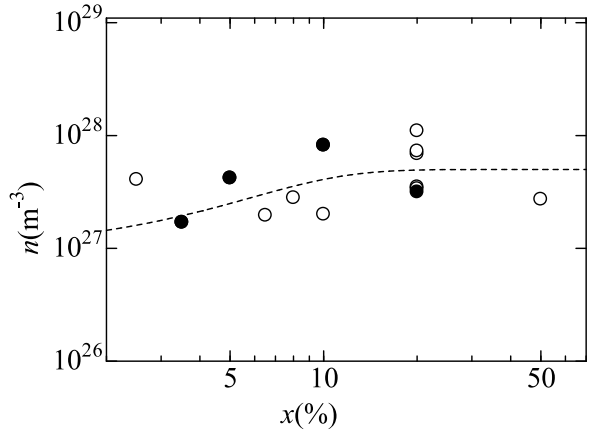}
\caption{\label{fig:density}
The m-cresol concentration-dependence of carrier density at room temperature. The temperature dependence of the films shown in the closed mark are shown in Fig.~\ref{fig:n_fit}. The dashed line provides a guide for the eyes.}
\end{figure}

At room temperature, the $x$-dependence of $n$ is shown in Fig.~\ref{fig:density}. The Hall measurement was limited to a narrow region of $x$, compared with that of the $\sigma$ measurement, due to a greater difficulty in accurate measurements of $\Delta V_\mathrm{y}$ in films of a higher resistance ($x\leq 1$ (\%)). At room temperature, the standard errors of mean (SEM) of the films $x$= 2.5, 3.5, 5(\%) were 7, 1.4, 0.6(\%) of the measured value of $n$, respectively. The reason for the data scattering is not due to the low accuracy of the Hall voltage measurement, as shown in Fig.~\ref{fig:typical_VB}, but may be due to the uncontrollable properties of the films, as mentioned above. Although the value of $n$ shows a weak $x$-dependence with some scattering, it seems that $n$ is approximately constant in the order of $\sim 10^{27}$/m$^3$. 
This value indicates that the carriers from the first dopant were fully excited at room temperature at any $x$ concentration.
\begin{figure}
\includegraphics[width=80mm]{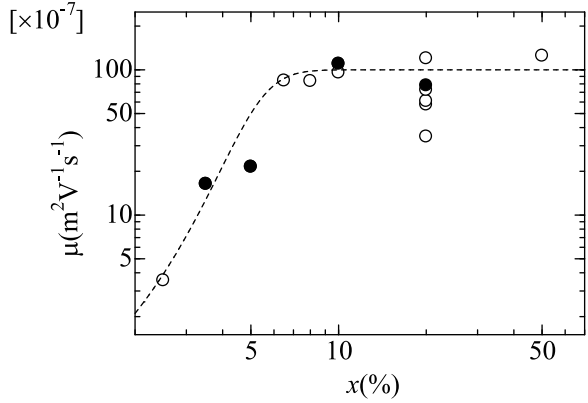}
\caption{\label{fig:mobility} 
The m-cresol concentration $x$-dependence of the mobility $\mu$ at room temperature. As $x$ increases, $\mu$ increases. The temperature dependence of the films shown by the closed marks is shown in Fig.~\ref{fig:mb_fit}. The dashed line provides a guide for the eyes.}
\end{figure}
For $x<10\%$, as shown in Fig.~\ref{fig:mobility}, the value of $\mu$ estimated from the relation $\mu = \sigma /(en) $ shows a similar $x$-dependence as that of $\sigma$. Considering the weak $x$-dependence of $n$, we may conclude that the $x$-dependence of $\sigma$ was mainly caused by the change in the mobility.

\begin{table*}
\caption{\label{table:free_electron} Quantities obtained using $m^\ast$, acquired in the last section at room temperature.}
  \begin{tabular}{cccccccc} 
    x(\%) & $\rho (\Omega \cdot \mathrm{m}) $ & $n(m^{-3})$ & $ \epsilon _F (eV) $ & $\ell(\mathrm{m})$ & $k_F\ell $ & $d \rho/dT(\Omega \cdot \mathrm{m/K}) $ \\
\hline
  3.5 & $ 2.2 \times 10^{-3} $ & $1.9 \times 10^{27} $ & 0.46 & $3.8 \times 10^{-12}$ & 0.014 & $-2.1 \times 10^{-6}$ \\ 
   5 & $6.8 \times 10^{-4} $ & $4.2 \times 10^{27} $ & 0.83 & $7.2 \times 10^{-12}$ & 0.036 & $5.1 \times 10^{-7}$ \\
  10 & $6.8 \times 10^{-5} $ & $8.4 \times 10^{27} $ & 1.44 & $4.6 \times 10^{-11}$ & 0.29 & $4.3 \times 10^{-3}$ \\ 
  20 & $2.5 \times 10^{-4} $ & $3.1 \times 10^{27} $ & 0.74 & $2.3 \times 10^{-11}$ & 0.11 & $1.1 \times 10^{-7}$ \\  
  \end{tabular} 
\end{table*}

Fig.~\ref{fig:resisivity-temp} (a) shows $\sigma (T)$ for the four films with different values of $x$, as denoted by the closed mark in Fig.~\ref{fig:conductivity}. The $\sigma (T) $ for all films shows the insulating characteristic $d\sigma/dT>0$ in the low temperature region, although the films with $x$ = 5\% and 10\% show metallic characteristics $d\sigma/dT<0$ at temperatures of $T>250$~K, as shown in Figs.~\ref{fig:resisivity-temp} (b) and (c). We will analyze these four films in detail. Before discussions on the $\sigma (T) $ characteristics, we will refer to the transport properties estimated from the free electron model: $\mu=e \tau /m^\ast$, $\ell=v_\mathrm{F} \tau$, $v_\mathrm{F}=\hbar k_\mathrm{F}/m^\ast$, and $k_\mathrm{F}=(3\pi^2 n)^{1/3}$, where $v_\mathrm{F}$ and $k_\mathrm{F}$ are the Fermi velocity and the Fermi wavenumber, respectively. Assuming the effective mass $m^\ast$ as that discussed in the below Eq. \eqref{eq:polaron_mob}, mean free path $\ell$, and Fermi energy $ \epsilon _F$ calculated at room temperature are listed in Table~\ref{table:free_electron}.
Even the longest $\ell \sim $ 0.046 nm of the film with $x =10$\% was shorter than the nearest distance of 0.14 nm between the carbon atoms. When considering the Ioffe-Regel criterion $k_\mathrm{F}\ell \sim 1$ for the metal-insulator (M-I) transition and the experimental fact that the quantities $k_\mathrm{F}\ell$ for the four films were in the range of $0.01<k_\mathrm{F}\ell<0.3$, the four films seemed to be in an insulating region. These results indicate that the transport properties of the preset films are no longer understood by the free electron model and the mobility may be influenced by the Anderson-localization effect. However, the discussion on the M-I criterion has not settled, and the criterion values are not focused\cite{Makise12JAP}. When considering the complex structures of the presented polymer systems, it was necessary to examine the exact structures and transport properties at sufficiently low temperatures for a detailed discussion on the M-I transition.

\begin{figure}
\includegraphics[clip, width=70mm]{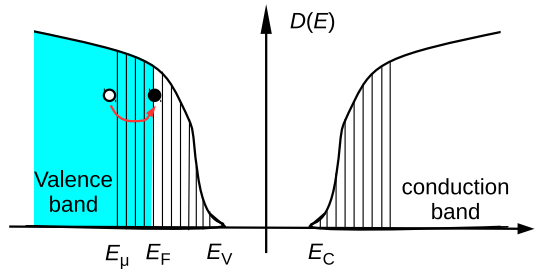}
\caption{\label{fig:bandgap} 
Schematics of the density of states, assumed by referring to three-dimensional amorphous inorganic materials.}
\end{figure}
 
The condition $\epsilon _\mathrm{F} \gg k_\mathrm{B}T$ based on the free electron model indicates the carrier fully degenerates and the carrier density is constant with temperature. However, the experimentally-obtained carrier density $n$ strongly depends on $T$, as shown in Fig.~\ref{fig:n_fit}.
Therefore, we assume the following Eq.~\eqref{eq:density_activation}:

\begin{equation}
 n \propto \exp \left( -\frac{\epsilon_n}{k_B T} \right). \label{eq:density_activation}
\end{equation}

After fitting Eq.~\eqref{eq:density_activation} to the experimental data, we obtained the carrier activation energy $\epsilon_n$ = 1.7, 3.5, 6.9, and 5.0 (meV) for the films with $x$ =3.5(\%), 5(\%), 10(\%), and 20(\%), respectively.
We obtained low-precision data of the temperature-dependence of $n$ of the sample with 3.5\% and could not obtain data of a sample below 3.5\%. Referring to the three-dimensional (3D) amorphous inorganic materials, we assumed the density of states of PANI, as shown in Fig~\ref{fig:bandgap}.
The bandgap of PANI ($\epsilon_\mathrm{g}=E_\mathrm{C}-E_\mathrm{V}$) was obtained experimentally in\cite{Stafstrom87PRL} ($\epsilon_\mathrm{g} \sim $1.5 eV) and theoretically in\cite{Kwon00JPC} ($\epsilon_\mathrm{g} \sim $1.3 eV), the values of which are larger than the $\epsilon_n$ obtained for the films used in this study.
This indicates that the simple band-conduction model cannot yet be applied but the localized states due to the Anderson localization effect are required in such dirty systems as the presented polymers.

A mobility edge is defined as the energy separating the localized and nonlocalized states in the conduction or valence bands of a 3D noncrystalline material. The hatched area by the vertical lines in Fig.~\ref{fig:bandgap} shows the localized discontinuous states. The blue area shows the electrons fully occupying the states. Unge and Christen determined the electron and hole mobility edges in amorphous polyethylene by calculating the boundary between the localized and delocalized states in energy-space\cite{Unge14CPL}.
When it is assumed that the electron excites with the activation energy $\epsilon_\mathrm{n}=E_\mathrm{F}-E_\mu$ and the generated hole under the hole mobility edge $E_\mu$ can move, the relation $n \propto e^{-\epsilon_n/k_BT}$ is expected. 
As $x$ increased, and the disorder decreased, it was predicted that $E_\mu$ increases, to result in the decreasing of $\epsilon_\mathrm{n}$ and increasing of $n$.
However, the experimental results show that $n$ increased but $\epsilon_n$ also increased with increasing $x$. This discrepancy suggests that the mobility edge cannot be sharply defined in the conducting polymer PANI.
For a distinct discussion, the Hall measurement is needed in the lower temperature region. There is a polaron mechanism for lowering mobility and localizing electrons. The carrier of the conducting polymers is a polaron-soliton, which has a solitary wave. The localization length can be much larger than the soliton size\cite{Sacha09PRL}. A polaron-soliton can be localized by Anderson localization.

Figs~\ref{fig:mb_fit} show the $T$-dependence of $\mu$ for the four films shown in Fig.~\ref{fig:resisivity-temp} and Fig.~\ref{fig:n_fit}. Although $\mu (T)$ for the films with $x = 10$\% and 3.5\% show a monotonic change, the $\mu (T)$ for the films with $x =5$\% show a maximum at around $T=250$~K. Such $T$-dependence, showing the transition between a positive and negative temperature coefficient of resistivity, has been reported for PANI blends \cite{Long04PHB} with $\sigma\sim10^4~\mathrm{S/m}$, although high-conductivity PANI films with $\sigma\sim10^5~\mathrm{S/m}$ can be successfully described by the simple Drude model without incorporating contributions from the disorder-induced localization theory\cite{Lee00nat}. For PANI-CSA and PANI-DBSA composite films, with the low conductivity mentioned above, the authors \cite{Long04PHB} analyzed $\rho(T)$ by the sum of the mechanism of quasi-1D metallic conduction in the ordered regions, and the 3D variable-range hopping conduction in the disordered region. Considering $\sigma(T)$ with $\sigma<10^4\mathrm{S/m}$ at room temperature, the contributions from the disorder-induced localization cannot be neglected.

In order to reproduce $\mu (T)$ for the present PANI films by the combination of opposite characteristics, we show some scattering mechanisms.
In ordered metallic polymers, the $\mu_\mathrm{ph}$ due to back-scattering thermal phonons is given by\cite{Heegerbook}:
\begin{equation}
\mu_\mathrm{ph}=\frac{ea^2m\omega_0t^2_0}{4\pi \alpha_1^2 \hbar^2} \exp \left( \frac{\hbar \omega_0}{k_B T} \right) \label{eq:phonon_mob}
\end{equation}
where $\alpha_1$ (eV/\AA) describes the magnitude of electron-phonon interaction, $\omega_0$ is the phonon frequency, $t_0$ is the transfer interaction (2.7 eV), and $a$ is the carbon-carbon spacing (0.14 nm).
In semiconducting crystals, ion-dopant atoms cause ionized impurity scattering\cite{Seegerbook}. Brooks and Herring calculated the ionized impurity scattering time as $\tau_\mathrm{ion} \sim T^{3/2}$. Using a characteristic temperature $T_1$ and the characteristic mobility ($\mu^\ast = 2.2 \times 10^{-5} ~(\mathrm{m}^2/\mathrm{Vs})$), we assume the following equation:

\begin{equation}
  \mu_\mathrm{ion}= \mu^\ast(T_1/T)^{-3/2} \label{eq:ion_mob}              
\end{equation}

Assuming that the total relaxation rate $1/\tau$ is expressed by the relationship: $1/\tau=1/\tau_\mathrm{ph}+1/\tau_\mathrm{ion}$, we attempted to fit the following expression:
\begin{equation}
  1/\mu=1/\mu_\mathrm{ph}+1/\mu_\mathrm{ion} \label{eq:total_mob}
\end{equation}
to the experimental data of the mobility. Using the fitting parameters $T_1, \omega_0$ and $\alpha_1 $ in Eqs.~\eqref{eq:phonon_mob} and \eqref{eq:ion_mob}, the black dashed lines in Fig.~\ref{fig:mb_fit} were calculated. We used the fitting parameters $T_1= 250,140,20, 53$ (K), $\alpha_1 = 0.135, 0.125, 0.235, 0.227$, and $\hbar\omega_\mathrm{0} = 0.017,  0.030 , 0.017, 0.024 $(meV) for the films with $x=3.5, 5, 10, 20$ (\%), respectively.

\begin{figure}
\includegraphics[width=120mm]{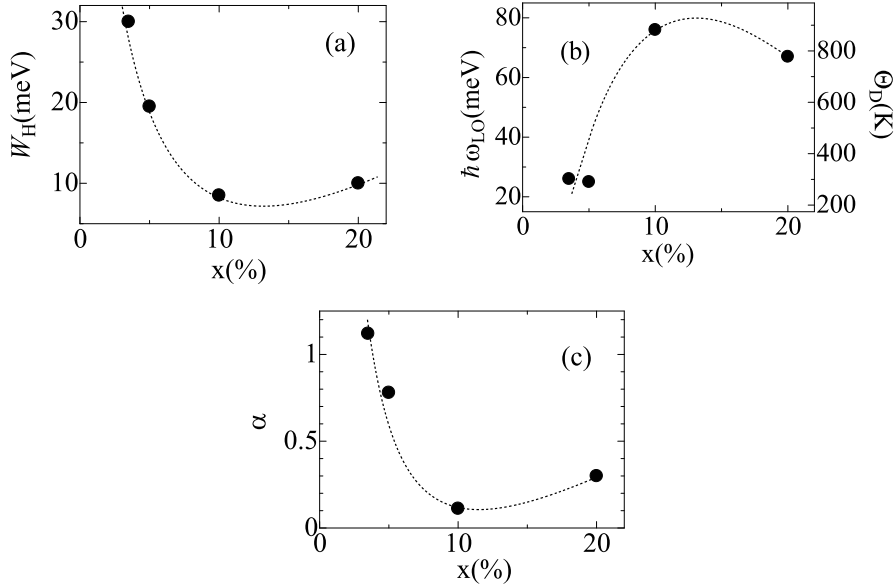}
\caption{\label{fig:ch_temp} 
The m-cresol concentration-dependence of (a) the hopping energy $W_\mathrm{H}$ and (b) the phonon energy $\hbar\omega_\mathrm{LO}$, obtained by fitting Eq.~\eqref{eq:polaron_mob} to the mobility $\mu (T)$ data, and (c) the electron-phonon coupling constant $\alpha$.
}
\end{figure}

\begin{table}
  \begin{tabular}{lcrrrrrr}
$x$(\%) & $\epsilon_\mathrm{n}$(meV) & $W_\mathrm{H}$(meV) & $\hbar \omega_\mathrm{LO}$(meV) \\
\hline
    3.5&  1.7 & 30 & 26\\ 
    5 & 3.5 & 19.5 & 25\\
    10&  6.9 & 8.5 & 76\\
    20&   5.0 & 10 & 67\\
  \end{tabular}
\caption{\label{table:ch_temp}The activation energy of carriers $\epsilon_\mathrm{n}$ obtained by fitting Eq.~(\ref{eq:density_activation}) to the $T$ dependence of the carrier density. The values $W_\mathrm{H}$ and $\hbar \omega_\mathrm{LO}$ were obtained by fitting Eq.~\eqref{eq:polaron_mob} to the $T$ dependence of the mobility.
} 
\end{table}

In contrast, the theory for the small polaron mobility \cite{Gurevich63SP, Devreese96EAP, Mottbook} results in the following formula:
 \begin{equation}
 \mu_\mathrm{sp}=\frac{ea^2\omega_\mathrm{LO}}{6k_\mathrm{B}T} \exp \left( -\frac{W_\mathrm{H}}{k_\mathrm{B}T} \right).  \label{eq:polaron_mob}
\end{equation}
where $a$ is the lattice constant of the crystal in which the small polaron occurs, $W_\mathrm{H}$ is the thermal activation energy for hopping and is given by half the small-polaron binding energy $E_\mathrm{b}$, and $\omega_\mathrm{LO}$ is the frequency of the longitudinal optical phonon. For $a$, we adopted the distance of carbon-carbon atoms as 0.14 nm. We fitted Eq.~\eqref{eq:polaron_mob} to the data of $\mu (T)$ using the fitting parameters $\omega_\mathrm{LO}$ and $W_\mathrm{H}$, as shown by the red lines in Fig.~\ref{fig:mb_fit}. The $x$-dependence of $W_\mathrm{H}$ and $\hbar \omega_\mathrm{LO}$ shown in Fig.~\ref{fig:ch_temp} (a) and (b) are reasonable, when it is taken into account that the increase in $x$ results in an increase of crystallinity.
These results suggest that the polymer becomes harder and the lattice deformation due to the polaron becomes weaker with the increase in $x$ ($x<10$). From the data of thermal conductivity, Nath {\it et al.} \cite{Nath13APL} estimated $\Theta_\mathrm{D}$ of PANI nanofibers without a secondary dopant to be approximately 240-300 K, which corresponds to $\hbar \omega_\mathrm{LO}$ = 21--26 meV under the assumption $\hbar \omega_\mathrm{LO} = k_\mathrm{B} \Theta _\mathrm{D}$. For metallic polyacetylenes with high conductivity $\sigma \sim 10^7$ (S/m), $\hbar \omega_0$ is 0.12 eV (1400 K) \cite{Heegerbook}.
These facts are consistent with the presented experimental result that $\hbar \omega_\mathrm{LO}$ increases with the increase in $\sigma$, that is, $x$.
Contrarily, MacDarmid and Epstein reported the absorption spectrum of PANI with various secondary dopants\cite{MacDiarmid94SM}. The absorption peak of a localized polaron was observed in the spectrum of PANI ($\sigma \sim 10^{-3}\mathrm{ S/m}$) but the peak diminished in that of PANI ($\sigma \sim 1.2 \times 10^4 \mathrm{S/m}$). This corresponds to the decrease in $W_\mathrm{H}$ with increasing $\sigma$, that is, $x$ in Fig.~\ref{fig:ch_temp}(a).

Strictly speaking, Eq.\eqref{eq:polaron_mob} is valid under the condition $T>\Theta_\mathrm{D}/2$. At the low temperature region, the contribution of quantum tunneling cannot be neglected\cite{Mottbook}. Then, it can be predicted that the mobility is larger than that estimated from Eq.\eqref{eq:polaron_mob}\cite{Holstein59AN}. This contribution may be the reason for the difference between the experimental data and the theory from Eq.~\eqref{eq:polaron_mob} in Fig.~\ref{fig:mb_fit}.

Feynman calculated the ground state energy $E_0$ of a polaron with the energy of an uncoupled electron-phonon system as zero energy and the mass $m^\ast$ of polaron as follows\cite{Feynman55PR}:
$E_0/(\hbar \omega _\mathrm{LO}) = -\alpha -0.0123\alpha^2-0.00064\alpha^3\cdots$ and $m^\ast/m_e=1+\alpha/6+0.025\alpha^2+\cdots$, where $\alpha$ is the electron-phonon coupling constant. Assuming that $E_0$ is equal to the polaron binding energy $-E_b =-2W_\mathrm{H}$, we obtained an $x$-dependence of $\alpha$, as shown in Fig.~\ref{fig:ch_temp}(c), then the effective mass ratio $m^\ast /m_\mathrm{e} = $1.48, 1.31, 1.04, 1.05 in the films with $x$=3.5, 5, 10, and 20, respectively. Using the value of $m^\ast$, we estimated the value of $\tau$, $E_\mathrm{F}$, and $T_\mathrm{F}$, given in Tables~\ref{table:free_electron} and \ref{table:ch_temp}. The increase in $x$ results in the decrease in the electron-phonon coupling strength. This result suggests that weakening of the electron-phonon coupling is needed for the enhancement of $\sigma$ of organic materials.

The polaron model fitted the data better than the band scattering model. The values of the parameters obtained by the fitting are reasonable. The Ioffe--Regel criterion $k_F\ell<1$ suggests that the hopping model is better than the band model. Thus, we conclude the polaron hopping model is suitable.
\section{Conclusions}
We measured the Hall effect of conducting polymer PANI films to obtain separately the mobility $\mu$ and the carrier density $n$ for PANI films with various concentrations $x$ of MCR. From these results, we found that the enhancement in $\sigma$ is mainly as a result of the change in the mobility $\mu$. We also measured the temperature-dependence of $n$ and $\mu$. With a decrease in $T$, $n$ decreases. We discussed the $x$-dependence of the activation energy of the carrier and concluded that a mobility edge cannot be clearly defined. We analyzed the $\mu(T)$ for four typical films with different values of $x$ using two scenarios: the combination of scattering models with opposite characteristics of $\mu(T)$ and the small-polaron hopping model.
The polaron model fitted the data better than the band scattering model. The parameters obtained by the fitting of the polaron model were reasonable.
With an increase in $x$, $W_\mathrm{H}$ and the electron-phonon coupling constant $\alpha$ decrease but the phonon energy $\omega_\mathrm{LO}$ increases. It is considered that these dependences cause enhancement by the crystallinity. Finally, the presented analysis suggests that the weakening of electron-phonon coupling is necessary to enhance the conductivity of organic materials.
\section{Acknowledgement}
The experiment was performed at the cryogenic center of Kyushu University. The authors thank Mr. Y. Ueda for the supply of liquid nitrogen. The authors thank Prof. N. Kuramoto for the preparation method of PANI. We thank Dr. K. Makise, Dr. N. Kokubo, and Dr. T. Asano for various contributions. English proofreading and publication fee were supported by JSPS KAKENHI Grant Number JP17K06303 and JP16K04877.

\section*{References}

\bibliography{PANI_HALL.bib}

\begin{thebibliography}{10}
\expandafter\ifx\csname url\endcsname\relax
  \def\url#1{\texttt{#1}}\fi
\expandafter\ifx\csname urlprefix\endcsname\relax\def\urlprefix{URL }\fi
\expandafter\ifx\csname href\endcsname\relax
  \def\href#1#2{#2} \def\path#1{#1}\fi

\bibitem{Heegerbook}
A.~Heeger, N.~S. Sariciftci, E.~B. Namdas, Semiconducting and Metallic
  Polymers, Oxford university Press, 2010.

\bibitem{Yamagishi10PRB}
M.~Yamagishi, J.~Soeda, T.~Uemura, Y.~Okada, Y.~Takatsuki, T.~Nishikawa,
  Y.~Nakazawa, I.~Doi, K.~Takimiya, J.~Takeya, Phys. Rev. B 81 (2010)
  161306(R).
\newblock \href {http://dx.doi.org/10.1103/PhysRevB.81.161306}
  {\path{doi:10.1103/PhysRevB.81.161306}}.

\bibitem{Uemura12PRB}
T.~Uemura, M.~Yamagishi, J.~Soeda, Y.~Takatsuki, Y.~Okada, Y.~Nakazawa,
  J.~Takeya, Temperature dependence of the hall effect in pentacene
  field-effect transistors: Possibility of charge decoherence induced by
  molecular fluctuations, Phys. Rev. B 85 (2012) 035313.
\newblock \href {http://dx.doi.org/10.1103/PhysRevB.85.035313}
  {\path{doi:10.1103/PhysRevB.85.035313}}.

\bibitem{Xinge11PRB}
X.~Yu, J.~Yu, J.~Zhou, J.~Huang, Y.~Jiang, Appl. Phys. Lett. 99 (2011) 063306.
\newblock \href {http://dx.doi.org/10.1063/1.3624586}
  {\path{doi:10.1063/1.3624586}}.

\bibitem{Lee12NM}
B.~Lee, Y.~Chen, D.~Fu, H.~T. Yi, K.~Czelen, H.~Najafov, V.~Podzorov, Trap
  healing and ultralow-noise hall effect at the surface of organic
  semiconductors, Nature Mater. 12 (2013) 1125--1129.
\newblock \href {http://dx.doi.org/10.1038/nmat3781}
  {\path{doi:10.1038/nmat3781}}.

\bibitem{Senanayak15PRB}
S.~P. Senanayak, A.~Z. Ashar, S.~P. Catherine~Kanimozhi, K.~S. Narayan,
  Room-temperature bandlike transport and hall effect in a high-mobility
  ambipolar polymer, Phys. Rev. B 91 (2015) 115302.
\newblock \href {http://dx.doi.org/10.1103/PhysRevB.91.115302}
  {\path{doi:10.1103/PhysRevB.91.115302}}.

\bibitem{Long04PHB}
Y.~Long, Z.~Chen, N.~Wang, J.~Li, M.~Wan, Electronic transport in
  {PANI-CSA/PANI-DBSA} polyblends, Physica B 344 (2004) 82--87.
\newblock \href {http://dx.doi.org/10.1016/j.physb.2003.09.245}
  {\path{doi:10.1016/j.physb.2003.09.245}}.

\bibitem{Lee00nat}
K.~Lee, S.~Cho, S.~H. Park, A.~J. Heeger, C.-W. Lee, S.-H. Lee, Metallic
  transport in polyaniline, Nature 441 (2006) 65--68.
\newblock \href {http://dx.doi.org/10.1038/nature04705}
  {\path{doi:10.1038/nature04705}}.

\bibitem{Kuramoto97POL}
N.~Kuramoto, A.~Tomita, Aqueous polyaniline suspensions: Chemical oxidative
  polymerization of dodecylbenzene-sulfonic acid aniline salt, Polymer 38
  (1997) 3055--3058.
\newblock \href {http://dx.doi.org/10.1016/S0032-3861(96)00861-0}
  {\path{doi:10.1016/S0032-3861(96)00861-0}}.

\bibitem{Kuramoto00SM}
S.-J. Su, N.~Kuramoto, Synthesis of processable polyaniline complexed with
  anionic surfactant and its conducting blends in aqueous and organic system,
  Synth. Met. 108 (2000) 121--126.
\newblock \href {http://dx.doi.org/10.1016/S0379-6779(99)00185-X}
  {\path{doi:10.1016/S0379-6779(99)00185-X}}.

\bibitem{MacDiarmid95SM}
A.~G. MacDiarmid, A.~J. Epstein, Secondary doping in polyaniline, Synth. Met.
  69 (1995) 85--92.
\newblock \href {http://dx.doi.org/10.1016/0379-6779(94)02374-8}
  {\path{doi:10.1016/0379-6779(94)02374-8}}.

\bibitem{MacDiarmid94SM}
A.~G. MacDiarmid, A.~J. Epstein, The concept of secondary doping as applied to
  polyaniline, Synth. Met. 65 (1994) 103--116.
\newblock \href {http://dx.doi.org/10.1016/0379-6779(94)90171-6}
  {\path{doi:10.1016/0379-6779(94)90171-6}}.

\bibitem{Stafstrom87PRL}
S.~Stafstr{\"o}m, J.~L. Bredas, A.~J. Epstein, H.~S. Woo, D.~B. Tanner, W.~S.
  Huang, A.~G. MacDiarmid, Polaron lattice in highly conducting polyaniline:
  Theoretical and optical studies, Phys. Rev. Lett. 59~(13) (1987) 1464--1467.
\newblock \href {http://dx.doi.org/10.1103/PhysRevLett.59.1464}
  {\path{doi:10.1103/PhysRevLett.59.1464}}.

\bibitem{Kohlman96PRL}
R.~S. Kohlman, J.~Joo, Y.~G. Min, A.~G. MacDiarmid, A.~J. Epstein, Phys. Rev.
  Lett. 77~(3) (1996) 2766--2769.
\newblock \href {http://dx.doi.org/10.1016/S0379-6779(96)04115-X}
  {\path{doi:10.1016/S0379-6779(96)04115-X}}.

\bibitem{Yamada12JP}
K.~Yamada, B.~Shinozaki, K.~Yano, H.~Nakamura, J. Phys. Conf. Ser. 400 (2012)
  042069.
\newblock \href {http://dx.doi.org/10.1088/1742-6596/400/4/042069}
  {\path{doi:10.1088/1742-6596/400/4/042069}}.

\bibitem{Varm12PI}
S.~J. Varm, F.~Xavier, S.~Varghese, S.~Jayalekshmi, Synthesis and studies on
  exceptionally crystalline polyaniline thin films, Polym Int 61 (2012)
  743--748.
\newblock \href {http://dx.doi.org/10.1002/pi.4131}
  {\path{doi:10.1002/pi.4131}}.

\bibitem{Makise12JAP}
K.~Makise, B.~Shinozaki, T.~Asano, K.~Mitsuishi, K.~Yano, K.~Inoue,
  H.~Nakamura, Relationship between variable range hopping transport and
  carrier density of amorphous {I}n2{O}3-10wt \% {Z}n{O} thin films, J. Appl.
  Phys. 112 (2012) 033716.
\newblock \href {http://dx.doi.org/10.1063/1.4745055}
  {\path{doi:10.1063/1.4745055}}.

\bibitem{Kwon00JPC}
O.~Kwon, M.~L. McKee, Calculations of band gaps in polyaniline from theoretical
  studies of oligomers, J. Phys. Chem. B 104 (2000) 1686--1694.
\newblock \href {http://dx.doi.org/10.1021/jp9910946}
  {\path{doi:10.1021/jp9910946}}.

\bibitem{Unge14CPL}
M.~Unge, T.~Christen, Electron and hole mobility edges in polyethylene from
  materialsimulations, Chem. Phys. Lett. 613 (2014) 15--18.
\newblock \href {http://dx.doi.org/10.1016/j.cplett.2014.08.058}
  {\path{doi:10.1016/j.cplett.2014.08.058}}.

\bibitem{Sacha09PRL}
K.~Sacha, C.~A. M{\"u}ller, D.~Delande, J.~Zakrzewski, Phys. Rev. Lett. 103
  (2009) 210402.
\newblock \href {http://dx.doi.org/10.1103/PhysRevLett.103.210402}
  {\path{doi:10.1103/PhysRevLett.103.210402}}.

\bibitem{Seegerbook}
K.~Seeger, Semiconductor Physics, Springer-Verlag, 2004.

\bibitem{Gurevich63SP}
V.~L. Gurevich, I.~G. Lang, Y.~A. Firsov, Sov. Phys. Solid State 4 (1963)
  918--.

\bibitem{Devreese96EAP}
J.~T. Devreese, Polarons, Encyclopedia of Applied Physics 14 (1996) 383--409.
\newblock \href {http://arxiv.org/abs/cond-mat/0004497v2}
  {\path{arXiv:cond-mat/0004497v2}}.

\bibitem{Mottbook}
S.~N. Mott, Conduction in Non-Crystalline Materials, Oxford university Press,
  1987.

\bibitem{Nath13APL}
C.~Nath, A.~Kumar, K.~Z. Syu, Y.~K. Kuo, Heat conduction in conducting
  polyaniline nanofibers, Appl. Phys. Lett. 103 (2013) 121905.
\newblock \href {http://dx.doi.org/10.1063/1.4821656}
  {\path{doi:10.1063/1.4821656}}.

\bibitem{Holstein59AN}
T.~Holstein, Studies of polaron motion: Part ii. the small polaron, Ann. Phys.
  8 (1959) 343--389.
\newblock \href {http://dx.doi.org/10.1016/0003-4916(59)90003-X}
  {\path{doi:10.1016/0003-4916(59)90003-X}}.

\bibitem{Feynman55PR}
R.~P. Feynman, Slow electrons in a polar crystal, Phys. Rev. 97 (1955)
  660--665.
\newblock \href {http://dx.doi.org/10.1103/PhysRev.97.660}
  {\path{doi:10.1103/PhysRev.97.660}}.

\end{thebibliography}

\end{document}